%% file: NPP.tex
\documentclass[aps,prd,twocolumn,showpacs,nofootinbib,floatfix]{revtex4-1}

\usepackage{preprintcover}  
\PreprintCoverPaperTitle{Search for non-pointing photons in the  diphoton and \MET \ \\ 
final state in $\sqrt{s}=\unit[7]{TeV}$ proton--proton collisions \\ using the ATLAS detector}  
\PreprintIdNumber{CERN-PH-EP-2013-049}  
\PreprintCoverAbstract{A search has been performed for photons originating in the decay of a neutral long-lived particle, 
exploiting the capabilities of the ATLAS electromagnetic calorimeter to make precise 
measurements of the flight direction of photons, as well as the calorimeter's excellent time resolution.
The search  has been made in the diphoton plus missing transverse energy final state,
using the full  data sample of 4.8~\ifb \ of 7~TeV proton--proton collisions collected in 2011 with the
ATLAS detector at the LHC. No excess is observed above
 the background expected from Standard Model processes. The results are used to set exclusion limits
in the context of Gauge Mediated Supersymmetry Breaking models, with the lightest neutralino being the
next-to-lightest supersymmetric particle and decaying 
with a lifetime in excess of $0.25$~ns into a photon and a gravitino. } 
\PreprintJournalName{Phys. Rev. D}  

\usepackage{atlasphysics} 
\usepackage{subfigure}
\usepackage{mathrsfs}
\usepackage{units}
\usepackage{graphics}
\usepackage{graphicx}
\usepackage{url}
\usepackage{xspace}
\usepackage{hyperref}
    
\newcommand{\gravitino}{\ensuremath{\tilde{G}}\xspace}

\begin{document}

\title{Search for non-pointing photons in the  diphoton and \MET \ final state in \\
$\sqrt{s}=\unit[7]{TeV}$ proton--proton collisions using the ATLAS detector}

\author{The ATLAS Collaboration}

\begin{abstract}
A search has been performed for photons originating in the decay of a neutral long-lived particle, 
exploiting the capabilities of the ATLAS electromagnetic calorimeter to make precise 
measurements of the flight direction of photons, as well as the calorimeter's excellent time resolution.
The search  has been made in the diphoton plus missing transverse energy final state,
using the full  data sample of 4.8~\ifb \ of 7~TeV proton--proton collisions collected in 2011 with the
ATLAS detector at the LHC. No excess is observed above
 the background expected from Standard Model processes. The results are used to set exclusion limits
in the context of Gauge Mediated Supersymmetry Breaking models, with the lightest neutralino being the
next-to-lightest supersymmetric particle and decaying 
with a lifetime in excess of $0.25$~ns into a photon and a gravitino. 
   
\end{abstract}

\maketitle 

\section{Introduction}
\label{sec:intro}

Supersymmetry~(SUSY)~\cite{Miyazawa:1966, Ramond:1971gb, Golfand:1971iw, Neveu:1971rx, Neveu:1971iv, Gervais:1971ji, Volkov:1973ix, Wess:1973kz, Wess:1974tw},
 a theoretically well-motivated candidate for physics beyond the
Standard Model~(SM), predicts the existence of a new SUSY partner (sparticle) for each of the SM particles, 
with identical quantum numbers except differing by  half a unit of spin. In R-parity conserving SUSY 
models~\cite{Fayet:1976et,Fayet:1977yc,Farrar:1978xj,Fayet:1979sa,Dimopoulos:1981zb}]
, these sparticles could be produced 
in pairs in proton--proton ($pp$) collisions at the CERN Large Hadron Collider (LHC), and would decay in cascades 
involving other sparticles and SM particles until the lightest SUSY particle~(LSP), which is stable, is produced.
In Gauge Mediated Supersymmetry Breaking~(GMSB)
models~\cite{Dine:1981gu,AlvarezGaume:1981wy,Nappi:1982hm,Dine:1993yw, Dine:1994vc,Dine:1995ag}, 
the LSP is the gravitino (\gravitino). GMSB phenomenology is largely determined by the properties of the
next-to-lightest supersymmetric particle~(NLSP). 
The results of this analysis are presented in the context of the so-called Snowmass Points and Slopes parameter set 8
(SPS8)~\cite{Allanach:2002nj}, which describes a set of minimal GMSB models with the lightest
neutralino (\ninoone) as the NLSP.
In the SPS8 set of models, the effective scale 
of SUSY breaking, denoted $\Lambda$, is a free parameter.  In addition, the \ninoone \ proper decay length, $c\tau(\ninoone)$, 
is a free parameter of the theory.

For the range of SPS8 $\Lambda$ values relevant for this analysis,
SUSY production is dominated by electroweak pair production
of gauginos, and in particular of \ninotwo \chinoonepm
\ and \chinoonep \chinoonem \ pairs. In the SPS8 models, 
the dominant decay mode of the NLSP is $\ninoone \ra \gamma + \gravitino$.
Previous ATLAS analyses have assumed prompt NLSP decays, with $c\tau(\ninoone) < 0.1$~mm,
and therefore searched for an excess production of diphoton events that, due to the escaping
gravitinos, exhibit significant missing transverse momentum~(\MET).
The latest such ATLAS results~\cite{2012afa} use the full 2011 dataset and, within the context of SPS8 models, exclude 
values of $\Lambda < 196$~TeV, corresponding to $m(\ninoone) > 280$~GeV, at the 95\% confidence level (CL). 
The limits on SPS8 models are less stringent in the case of a longer-lived NLSP. For example, 
recent CMS 95\% CL limits~\cite{Chatrchyan:2012vxa}, obtained using the \MET \ spectrum of events
with at least three jets and one or two photons, coupled with measurements of the photon
arrival time,  require $m(\ninoone) > 220$~GeV for $c \tau(\ninoone)$ values up to 500~mm.

The analysis reported in this paper uses the full  data sample of 4.8~\ifb \ of 7~TeV $pp$ collisions collected in 2011 with the
ATLAS detector at the LHC, and considers the scenario where the \ninoone \ has a finite lifetime
and can travel some distance from its production point  before decaying.  
The search is performed in an inclusive sample of candidate
diphoton + \MET \ events. 

The long-lived NLSP scenario introduces the possibility of a decay photon being produced after a
finite delay and with a flight direction that does not point back to the primary vertex (PV) of the event. 
The analysis searches for such ``non-pointing photons'' by exploiting the fine segmentation
of the ATLAS electromagnetic (EM) calorimeter to measure the flight direction of photons.
The variable used as a measure of
the degree of non-pointing of the photon is $z_{\rm DCA}$, the difference between the 
$z$-coordinate\footnote{ATLAS uses a 
right-handed coordinate system with its origin at the nominal interaction point (IP) 
in the center of the detector and the $z$-axis along the beam pipe. The $x$-axis 
points from the IP to the center of the LHC ring, and the $y$-axis points upward. 
Cylindrical coordinates $(r,\phi)$ are used in the transverse plane, $\phi$ being 
the azimuthal angle around the beam pipe. The pseudorapidity is defined in terms 
of the polar angle $\theta$ as $\eta=-\ln\tan(\theta/2)$.} 
of the photon extrapolated back to its distance-of-closest-approach (DCA) to the beamline (i.e. $x = y = 0$) 
and $z_{\rm PV}$, the $z$-coordinate of the PV.
The search for non-pointing photons is then performed by fitting
the shape of the $z_{\rm DCA}$ distribution obtained for photons in the signal region, defined as diphoton events with $\MET > 75$~GeV,
to a combination of templates that describe the $z_{\rm DCA}$ distribution for the expected signal and
background events. In addition, 
the excellent time resolution of the calorimeter is exploited to
measure the arrival times of the photons, providing a cross-check of the results.

\section{The ATLAS Detector}
\label{sec:detector}

 The ATLAS detector~\cite{Aad:2008zzm} covers nearly the entire solid 
 angle around the collision point, and consists of an inner tracking detector surrounded 
 by a solenoid, EM and hadronic calorimeters, 
 and a muon spectrometer incorporating three large toroidal magnet systems.
The ATLAS inner-detector system (ID) is immersed in a 2~T axial magnetic field, provided
by a thin superconducting solenoid located before the calorimeters, and 
provides charged particle tracking in the pseudorapidity range $|\eta| < 2.5$. 
The ID consists of three detector subsystems, beginning closest to the beamline
with the high-granularity silicon pixel detector, 
followed at larger radii by the silicon microstrip tracker and then the straw-tube-based
transition radiation tracker.  The ID allows an accurate reconstruction of tracks from the primary $pp$ 
collision and precise determination of the location of the PV.
The ID also identifies tracks from secondary vertices, permitting the efficient identification of photons
that convert to electron--positron pairs as they pass through the detector material. 

This analysis relies heavily on the capabilities of the ATLAS calorimeter system, which 
covers the pseudorapidity range $|\eta|< 4.9$. Finely-segmented
EM calorimetry is provided by barrel ($|\eta| < 1.475$) and end-cap ($1.375 < |\eta| < 3.2$) 
lead/liquid-argon (lead/LAr) EM sampling calorimeters. An additional thin LAr 
presampler covering $|\eta| < 1.8$ allows corrections for energy losses in material upstream of the EM calorimeters. 
Hadronic calorimetry is provided by a steel/scintillator-tile calorimeter, segmented into three barrel 
structures within $|\eta| < 1.7$, and two copper/LAr hadronic end-cap calorimeters. The solid 
angle coverage is completed with forward copper/LAr and tungsten/LAr calorimeter modules 
optimized for EM and hadronic measurements, respectively.  Outside the ATLAS calorimeters lies the
muon spectrometer, which identifies and measures the deflection of muons up to $|\eta| = 2.7$,
 in a magnetic field generated by superconducting air-core toroidal magnet systems.

 \subsection{Pointing resolution}
 
 The EM calorimeter is segmented into three layers in depth, that are used to measure the longitudinal profile
of the shower. The first layer uses highly
 granular ``strips'' segmented in the $\eta$ direction, designed to allow efficient discrimination between single 
 photon showers and two overlapping showers originating, for example, from the decay of a $\pi^0$ meson. The second layer
collects most of the energy deposited in the calorimeter by EM showers initiated by electrons or photons. 
Very high energy showers can leave significant energy deposits in the third layer, which can also
 be used to correct for energy leakage beyond the EM calorimeter.
By measuring precisely the centroids of the EM shower in the first and second layers,
the flight direction of photons can be determined. 
In the ATLAS $H \ra \gamma\gamma$ analysis~\cite{2012gk} that contributed to the discovery of a 
Higgs-like particle, this capability of the EM calorimeter was used to help choose the PV
from which the two photons originated, thereby improving the diphoton invariant mass resolution and
sensitivity of the search.  The analysis described in this paper uses the measurement of the
photon flight direction to search for photons that do not point back to the PV.
The angular resolution of the EM calorimeter's measurement of the flight direction of prompt photons
is about 60~mrad/$\sqrt{E}$, where $E$ is the energy measured in GeV. This angular precision
corresponds, in the EM barrel calorimeter, to a resolution on $z_{\rm DCA}$ of about 15~mm for prompt photons with
 energies in the range of 50--100~GeV. Given the geometry, the $z_{\rm DCA}$
resolution is worse for photons reconstructed in the end-cap calorimeters, so the pointing analysis is 
restricted to photon candidates in the EM barrel calorimeter.

While the geometry of the EM calorimeter has been optimized for detecting particles that point back to near 
the nominal interaction point at the center of the detector (i.e. $x = y = z = 0$), the fine segmentation allows good  
pointing performance to be achieved over a  wide range of photon impact angles.  
Figure~\ref{fig:pointing}  shows, as a function of $| z_{\rm DCA} |$, the expected pointing resolution
for SPS8 signal photons from Monte Carlo (MC) simulations, obtained by fitting to a Gaussian the difference between the value of $z_{\rm DCA}$
obtained from the calorimeter measurement and the MC generator-level information. 
The pointing resolution degrades with increasing $| z_{\rm DCA} |$, 
but remains much smaller than $| z_{\rm DCA} |$ in the region where the signal is expected.

The calorimeter pointing performance has been
verified in data by using the finite spread of the LHC collision region along the $z$-axis. Superimposed on
Figure~\ref{fig:pointing} is the pointing resolution achieved for a sample of electrons from $Z \ra ee$ events,
where the distance, $z_{\rm PV}$, between the PV and the nominal center of the detector
serves the role of $z_{\rm DCA}$. In this case, the pointing resolution is obtained 
by fitting to a Gaussian the difference between $z_{\rm PV}$, as determined with high precision using tracking information,
and the calorimeter measurement of the origin of the electron, along the beamline.
Figure~\ref{fig:pointing} shows that similar
pointing performance is observed for photons and for electrons, as expected given their similar EM shower developments.
This similarity validates the use of a sample of electrons from $Z \ra ee$ events to study the 
pointing performance for photons. The expected pointing performance for electrons in a MC sample of
$Z \ra ee$ events is also shown on Figure~\ref{fig:pointing}, and is consistent with the data.
  The level of agreement between MC simulation and data over the range of values that can be accessed 
in the data gives confidence in the extrapolation using MC simulation to the larger deviations from pointing characteristic
of the signal photons.

\begin{figure*}[hbpt]
\centering
\includegraphics[width=0.9\textwidth]{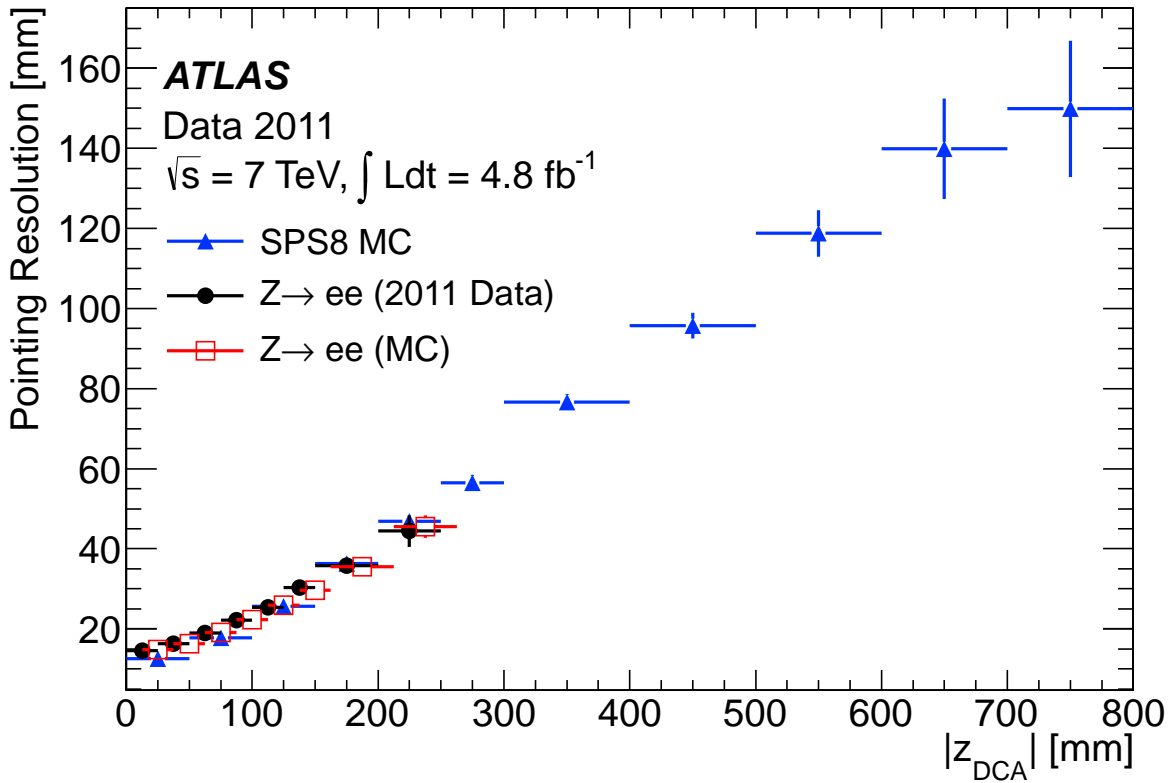}
\caption{Pointing resolution obtained for EM showers in the ATLAS LAr EM barrel calorimeter. 
The pointing resolution for photons from GMSB signal MC samples is plotted as a function of $| z_{\rm DCA} |$.
The pointing resolution is also shown for $\Zee$ \ data and MC, for which 
the primary vertex position, $z_{\rm PV}$, serves the role of $z_{\rm DCA}$. }
\label{fig:pointing}
\end{figure*}

\subsection{Timing resolution}
\label{sec:timing}

Photons from NLSP decays would reach the LAr calorimeter with
a slight delay compared to prompt photons. This 
delay results mostly from the flight time of the heavy NLSP, 
 as well as some effect due to the longer geometric path of
  a non-pointing photon produced in the NLSP decay.
The EM calorimeter, with its novel ``accordion'' design, and its readout, which
incorporates fast shaping, has excellent timing performance.
Quality control tests during production of
the electronics required the clock jitter on the LAr readout boards to be less than 20~ps, 
with typical values of 10~ps~\cite{Buchanan:2008zza}. Calibration tests of the overall
electronic readout performed {\sl in situ} in the ATLAS cavern show a timing resolution
of $\approx 70$~ps~\cite{1748-0221-5-09-P09003}, limited not by the readout but by the jitter of
the calibration pulse injection system. Test-beam measurements~\cite{2008NIMPA.597..178A} 
of production EM barrel calorimeter modules demonstrated a timing resolution
of $\approx 100$~ps in response to high energy electrons.

For this analysis, the arrival time of an
EM shower is measured using the second-layer EM calorimeter cell with the
maximum energy deposit. During 2011,
the various LAr channels were timed-in online with a precision of about 1~ns.   
A large sample of $W \rightarrow e\nu$  events was used to determine
several calibration corrections which were applied to further optimize the
timing resolution for EM clusters.  The calibration includes corrections of
various offsets in the timing of individual channels, corrections for the energy
dependence of the timing, and flight-path corrections depending on the
position of the PV. The corrections determined using the  $W \rightarrow e\nu$
events were subsequently applied to electron candidates in $Z  \rightarrow ee$ 
events to validate the procedure as well as to determine the timing
performance in an independent data sample. Figure~\ref{fig:timeresn} shows
the time resolution achieved as a function of the energy deposited in the
second-layer cell used in the time measurement. The time resolution, $\sigma(t)$, is expected to follow
the form $\sigma(t) = a/E \oplus b$, where $E$ is the energy measured in GeV and
$\oplus$ indicates addition in quadrature. Superimposed on Figure~\ref{fig:timeresn}
is the result of a fit of the resolution to this form, where the parameters
$a$ and $b$ multiply the so-called noise term and constant term, respectively.
A timing resolution of $\approx 290$~ps is achieved  for a large energy deposit. By comparing
the arrival times of the two electrons in $Z \rightarrow ee$ events, this resolution is
understood to include a correlated contribution of $\approx 220$~ps, as expected due
to the spread in $pp$ collision times caused by
the lengths of individual proton bunches along the LHC beamline. Subtracting this beam
contribution in quadrature, the obtained timing resolution for the calorimeter is
$\approx 190$~ps. 

\begin{figure}[hbpt]
\centering
\includegraphics[width=\columnwidth]{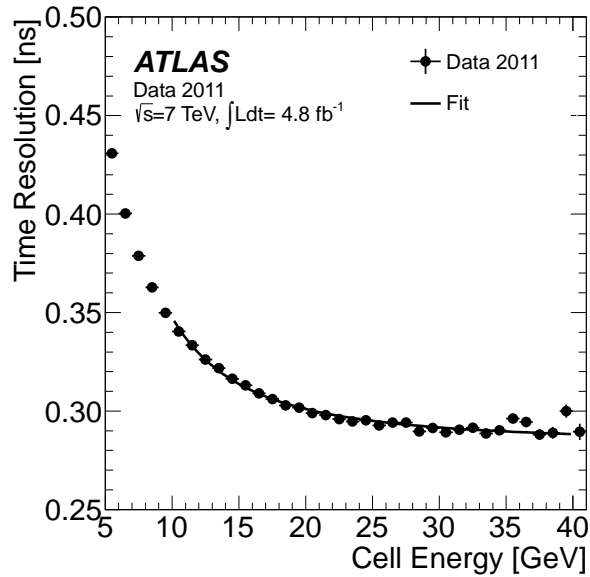}
\caption{Time resolution obtained for EM showers in the ATLAS LAr EM barrel calorimeter,
as a function of the energy deposited in the second-layer cell
with the maximum deposited energy.  Superimposed is the 
result of the fit described in the text. The data are shown for electrons read out using
high gain, and the errors shown are statistical only.}
\label{fig:timeresn}
\end{figure}

To cover the full dynamic range of physics signals of interest, the
ATLAS LAr calorimeter readout boards~\cite{Buchanan:2008zza} employ three
overlapping linear gain scales, dubbed high, medium, and low, where
the relative gain of successive scale is reduced by a factor of about ten.  For a given event,
any individual LAr readout channel is digitized using the gain scale that provides
optimal energy resolution, given the energy deposited in that calorimeter cell. 
The results in Figure~\ref{fig:timeresn} are those obtained for
electrons where the time was measured using a second-layer cell read out
using high gain, for which the $W \rightarrow e\nu$ sample used to calibrate the
timing is large. Calibration samples for the medium and low gain scales are smaller, resulting in reduced precision.  
The timing resolutions obtained are about 400~ps for medium gain and about 1~ns for low gain.

\section{Data and Monte Carlo Simulation Samples}
\label{sec:datamc}

This analysis uses a data sample of $pp$ collisions at a center-of-mass-energy of
$\sqrt{s}=\unit[7]{TeV}$, recorded with the ATLAS detector in 2011.  The
data sample, after applying quality criteria that require all ATLAS subdetector
systems to be functioning normally, corresponds to a total integrated
luminosity of $4.8 \pm 0.1$~fb$^{-1}$.

While all background studies, apart from some
cross-checks, are performed with data, MC simulations are
used to study the response to SPS8 GMSB signal models, as a function
of the free parameters $\Lambda$ and $c\tau$.  The other SPS8 model parameters are fixed to the following
values: the messenger mass $M_{\rm mess} =2\Lambda$, the number
of SU(5) messengers $N_5 = 1$, the ratio of the vacuum expectation
values of the two Higgs doublets $\tan \beta = 15$,
and the Higgs-sector mixing parameter $\mu > 0$~\cite{Allanach:2002nj}.
The SPS8 SUSY mass spectra, branching ratios and decay widths 
are calculated using {\sc ISAJET}~\cite{isajet} version~7.80. The
signal yield is normalized to the
central value of the GMSB signal cross section, as calculated at
next-to-leading order (NLO) using {\sc Prospino }~\cite{prospino} version~2.1
with the CTEQ6.6m~\cite{Nadolsky:2008zw} parton distribution functions (PDFs).  
The total uncertainty on the signal cross section, times the signal acceptance and  efficiency,   from varying the PDFs and 
factorization and renormalization scales,
as described in Ref.~\cite{Kramer:2012bx}, ranges from 4.7\% to 6.4\%, depending on $\Lambda$.

The Herwig++ generator version 2.4.2~\cite{Bahr:2008pv} 
with MRST\,2007\,LO${}^{*}$~\cite{Sherstnev:2007nd}  PDFs is used to generate the signal MC samples.
The branching ratio of the  $\ninoone \ra \gamma + \gravitino$ decay mode
is fixed to unity.
Signal samples were generated for fixed values of $\Lambda$, ranging from
70~TeV to 210~TeV in 10~TeV steps, and for a variety of values of the
NLSP lifetime. The performance for any lifetime can be obtained by reweighting
as a function of lifetime the existing samples for a given value of $\Lambda$. The
current analysis considers NLSP lifetime values in excess of $0.25$~ns.

All MC samples were processed with the  simulation~\cite{Aad:2010wq} of the ATLAS
detector based on {\sc GEANT4}~\cite{geant4} and were reconstructed with the same
algorithms used for the data. The presence of additional $pp$ interactions
(pile-up) as a function of the instantaneous luminosity is
taken into account by overlaying simulated minimum bias
events according to the distribution of the number
of pile-up interactions observed in data, with an average of about nine
interactions per bunch crossing. The MC samples were generated with a spatial 
distribution of the PV consistent with that observed in the data.
The simulated LAr time values are smeared such that the time resolution matches 
the performance, as described in Sec.~\ref{sec:timing}, observed in data.

\section{Object Reconstruction and Identification}
\label{sec:objects}

The reconstruction of converted and unconverted photons
and of electrons is described in Refs.~\cite{Aad:2010sp} and 
\cite{Aad:2011mk}, respectively.
Shape variables computed from the lateral and longitudinal
energy profiles of the EM shower in the calorimeter are
used to identify photons and discriminate against backgrounds. 
Two sets of selection criteria, denoted ``loose'' and ``tight'', are defined~\cite{Aad:2010sp}. 
The loose photon identification, designed for high photon
efficiency with modest background rejection, uses variables describing the
shower shape in the second layer of the EM calorimeter, as
well as leakage of energy into the hadronic calorimeter.
The tight photon identification, designed for higher purity
photon identification with still reasonable efficiency, includes more 
stringent cuts on these variables, as well as requirements on
additional variables describing the shower in the first layer
of the EM calorimeter.  The various selection
criteria do not depend on the transverse momentum of the photon (\ET), but do vary as a function of $\eta$
in order to take into account variations in the calorimeter geometry and
in the thickness of the upstream material. For more details,
see Ref.~\cite{Aad:2010sp}.

The measurement of \MET~\cite{Aad:2012re}  is
based on energy deposits in calorimeter cells inside three-dimensional
clusters with $|\eta| < 4.5$.
The energy of each cluster is calibrated to correct for the different response
to electromagnetically and hadronically induced showers,
energy loss in dead material, and out-of-cluster energy. The value of
\MET \  is corrected for contributions from any muons identified in the event, 
by adding in the energy derived from the properties of reconstructed muon tracks.

\section{Event Selection}
\label{sec:selection}

The selected events were collected by an online trigger requiring the presence 
 of at least two loose photon candidates, each with $\ET > 20$~GeV
and $|\eta| < 2.5$.  
To ensure the selected events resulted from a beam collision, 
events are required to have at least one PV
candidate with five or more associated tracks. In case of multiple
vertices, the PV is chosen as the 
vertex with the greatest sum of the square of the transverse momenta of all associated tracks.

The offline photon selection requires the two photon candidates to each have 
$\ET > 50$~GeV, and to satisfy $|\eta| < 2.37$, excluding the
transition region of $1.37 < |\eta| < 1.52$ between the barrel and end-cap EM calorimeters.
In addition, both photons are required to be isolated: after correcting for contributions 
from pile-up and the deposition ascribed to the photon itself~\cite{Aad:2010sp}, the transverse
energy deposited in the calorimeter in a cone of $\Delta R = \sqrt{(\Delta \eta)^2 + (\Delta \phi)^2}  = 0.2$
around each photon candidate must be less than 5~GeV~\cite{Aad:2011zj}.

Due mostly to the inclusion of cuts on the EM shower shape in the very finely
segmented strips of the first EM calorimeter layer,
the efficiency of the tight photon identification decreases with $|z_{\rm DCA}|$ for
values of  $|z_{\rm DCA}|$ larger than $\approx 100$~mm. The loose efficiency
remains flat over a wider range of $|z_{\rm DCA}|$, up to values of $\approx 250$~mm,
after which it decreases less rapidly
than the tight efficiency. The event selection requires
at least one of the isolated photon candidates 
to pass the tight photon identification requirements, while the other
must pass the loose photon identification cuts. The
selected sample will therefore be referred to hereafter as the Tight--Loose (TL) diphoton sample.
To reduce the potential bias in the pointing
measurement that results from applying the photon identification requirements, only
the loose photon in each event is examined for evidence of non-pointing. 
The $\eta$ cut on the loose photon is tightened to restrict 
it to lie in the EM barrel calorimeter, namely $|\eta| < 1.37$.

The TL diphoton sample is divided into exclusive subsamples according
to the value of \MET. The TL sample with $\MET < 20$~GeV is used, as described in Sec.~\ref{sec:bkgnds},
to model the prompt backgrounds. The TL events with intermediate \MET \ values, 
namely $20 \ {\rm GeV} < \MET < 75$~GeV, 
are used as a control sample to validate the analysis procedure.
The final signal region, which contains a total of 46 selected events, 
is defined by applying to the TL diphoton sample the additional requirement that the
value of \MET \ exceeds $75$~GeV. 

Table~\ref{tab:effic} summarizes the total acceptance times efficiency of the
selection requirements, for examples of SPS8 signal model points with various values of $\Lambda$ and $\tau$.
For fixed $\Lambda$, the acceptance falls approximately exponentially with increasing $\tau$, dominated by the requirement
that both NLSPs decay inside the ATLAS tracking detector (which extends to a radius of 107~cm) so that the decay photons are detected
by the EM calorimeters. For fixed $\tau$, the acceptance increases
with increasing $\Lambda$,
since the SUSY particle masses increase, leading the decay cascades to produce, on average, higher  \MET \
and also higher \ET \ values of the decay photons.

\begin{table}[htbp]
\begin{center}  
\begin{tabular}{ccccccc}          \hline	  \hline
$\tau$   &  \multicolumn{3}{c}{$\Lambda$  (TeV)} \\ \cline{2-4}
 (ns) & 80          & 120                &  160  \\ \hline
\ \   0.25 \ \   & $15.3 \pm 0.3$ \ &  $29.6 \pm 0.3$ \ & $45.1 \pm 0.3$   \   \\ 
  1 & $11.1 \pm 0.1$ \ & $27.0 \pm 0.2$ \ & $35.9 \pm 0.3$ \  \\ 
  6 & $2.01 \pm 0.02$ & $5.38 \pm 0.02$ & $8.06 \pm 0.06$ \\ 
  20 & $0.39 \pm 0.01$ & $1.006 \pm 0.005$ & $1.43 \pm 0.01$ \\ 
  40 & $0.175 \pm 0.005$ & $0.384 \pm 0.002$ & $0.510 \pm0.004$ \\ 
  80 &\ $0.090 \pm 0.004$ \  & \ $0.164 \pm 0.001$ \ & \ $0.196 \pm0.002$ \  \\ \hline \hline
        \end{tabular}
    \caption{The total signal acceptance times efficiency, in percent, of the event selection requirements,
 for sample SPS8 model points with various $\Lambda$ and $\tau$ values. The uncertainties
 shown are statistical only.\label{tab:effic} }
  \end{center}
\end{table}

\section{Signal and Background Modeling}
\label{sec:templates}

\subsection{SPS8 GMSB signal}

The shape of the $z_{\rm DCA}$ distribution, alternately denoted hereafter as the pointing distribution, 
that is expected for photons from NLSP decays in events passing the selection cuts
is determined using the SPS8 GMSB MC signal samples described previously, for 
various values of $\Lambda$ and $\tau$.  The signal pointing distributions, normalized
to unit area, are used as signal templates, hereafter referred to as $T_{\rm sig}$.
Since the $T_{\rm sig}$ shape is determined using MC simulations, systematic uncertainties in the shape
are included to account for possible differences in pointing performance
between MC simulation and data. In particular, the presence of pile-up in the collisions could 
impact the pointing resolution, due both to energy deposits in the calorimeter from
additional minimum bias collisions and to the possibility of misidentifying the
PV.  These effects are modeled in the MC simulation. 
However, as a conservative estimate of the systematic $T_{\rm sig}$ shape variations that could 
occur due to pile-up, the differences are taken between the nominal shape and those from the 
two subsamples which are obtained by dividing the MC samples roughly in two according 
to the number of PV candidates identified in each event.
The $T_{\rm sig}$ distributions, along with their statistical and
total uncertainties, are shown in Figure~\ref{figJetZvertTemplates}
for $\Lambda = 120$~TeV and for NLSP lifetime values of $\tau = 0.5$~and~$30$~ns.

\begin{figure*}[hbpt]
\centering
\includegraphics[width=\textwidth]{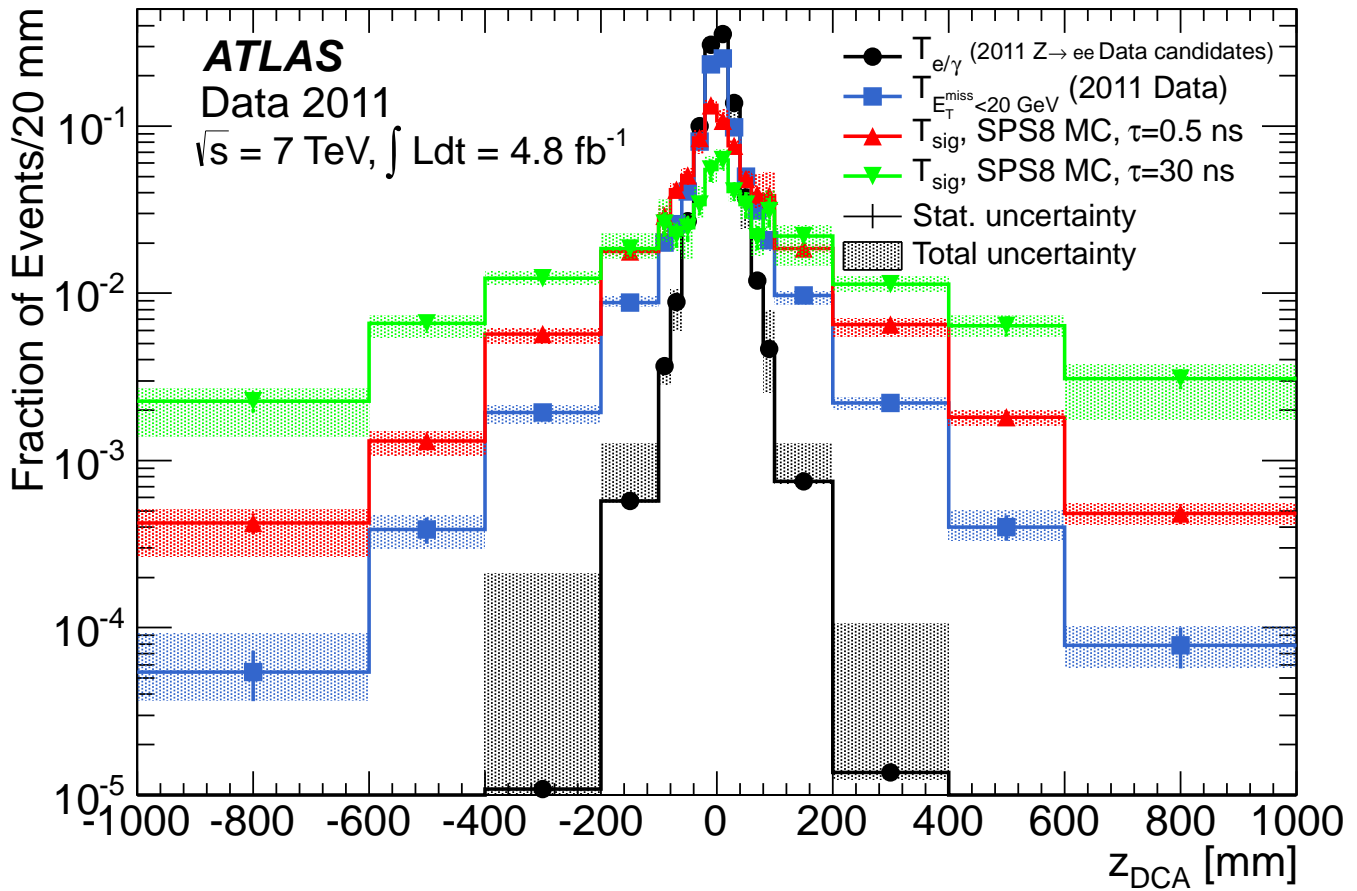}
\caption{The $z_{\rm DCA}$ templates from $Z \ra ee$ events, from the 
TL control sample with \MET \ less than 20~GeV, and for MC simulations
of GMSB signals with $\Lambda = 120$~TeV and values for the 
NLSP lifetime of $\tau = 0.5$~and~$30$~ns. 
The data points show the
statistical errors, while the shaded bands show the total uncertainties, with
statistical and systematic contributions added in quadrature. 
The first (last) bin includes the contribution from underflows (overflows).}
\label{figJetZvertTemplates}
\end{figure*}

\subsection{Backgrounds}
\label{sec:bkgnds}

The background is expected to be completely dominated
by $pp$ collision events, with possible backgrounds due to cosmic rays, beam-halo events, 
or other non-collision processes being negligible.
The source of the loose photon in background events contributing to the selected TL sample
is expected to be either a prompt photon, an electron misidentified as a photon, or a jet misidentified as a photon. In
each case, the object providing the loose photon signature originates from the PV. 
The pointing and timing distributions expected for these background sources are determined using
data control samples.

Given their similar EM shower developments, the pointing resolution is similar
for prompt photons and for electrons.  The $z_{\rm DCA}$ distribution expected for prompt photons 
and for electrons is therefore modeled using electrons in $Z\ra ee$ data events. 
The $Z\ra ee$ event selection requires a pair of oppositely charged electron candidates, each of which 
has $\pT > 25$~GeV and $|\eta| < 2.37$ (excluding the transition region between the barrel 
and end-cap calorimeters). One of the electron candidates, dubbed 
the ``tag'', is required to pass additional selection cuts on the EM shower topology and 
tracking information designed to identify a high-purity sample of electrons. The second electron candidate, 
dubbed the ``probe'', is restricted to the range $|\eta| < 1.37$. 
The dielectron invariant mass is required to agree with the value of the $Z$ mass within 10~GeV, 
a requirement that produces a sufficiently clean sample of $Z \ra ee$ events. To avoid any bias, 
the pointing resolution for electrons is determined using the distribution measured for the probe electrons. 
The pointing distribution determined from $Z\ra ee$ events, 
normalized to unit area, is used as the pointing template for prompt photons
and electrons, and is referred to hereafter as $T_{e/\gamma}$. 

While the EM showers of electrons and photons are similar, there are some differences.
In particular, electrons traversing  the material of the ID may emit bremsstrahlung photons,
widening the resulting EM shower. In addition, photons can convert into electron--positron
pairs in the material of the ID. In general, the EM showers of unconverted photons are
slightly narrower than those of electrons, which are in turn slightly narrower than those of converted
photons. The EM component of the background in the signal region includes a mixture of electrons, converted
photons, and unconverted photons.  Therefore, using electrons from $Z\ra ee$ events to model the EM showers of the loose
photon candidates in the signal region can slightly underestimate the pointing resolution in some cases, and slightly
overestimate it in others.  The pointing distribution from $Z\ra ee$ events is
taken as the nominal $T_{e/\gamma}$ shape. The distributions from MC samples of
unconverted and converted photons, with similar kinematic properties as expected for signal photons,
are separately taken to provide conservative estimates of the possible
variations in the $T_{e/\gamma}$ shape which could result from not
separating these various contributions. The $T_{e/\gamma}$ distribution, along with its statistical
and total uncertainties, is shown superimposed on Figure~\ref{figJetZvertTemplates}.

Due to their wider showers in the calorimeter,  jets have a
wider $z_{\rm DCA}$ distribution than prompt photons and electrons. The sample of events passing
the TL selection, but with the additional requirement that $\MET < 20~$GeV, is used as a data control sample that
includes jets with properties similar to the background contributions expected 
in the signal region. The \MET \ requirement  serves to render negligible any possible signal
contribution in this control sample. 
The shape of the $z_{\rm DCA}$ distribution for the loose photon in these events,
normalized to unit area, is used
as a template, referred to hereafter as $T_{\MET<20 \ {\rm GeV}}$,  in
the final fit to the signal region.  
The TL sample with $\MET < 20$~GeV should be
dominated by  jet--jet, jet--$\gamma$ and $\gamma\gamma$ events. 
Therefore, the $T_{\MET<20 \ {\rm GeV}}$ template includes contributions from photons as well
as from jets faking the loose photon signature.  When using the template in the fit to extract
the final results,  it is not necessary to separate the photon and jet
contributions.  Instead, the relative fraction of the two background templates is treated as a nuisance parameter
in the fitting procedure, as discussed in Sec.~\ref{sec:results}.

The pointing resolution depends on the value of \ET \ of the photon 
candidate.   Applying the shape of the $T_{\MET <20 \ {\rm GeV}}$ template to describe 
events in the signal region, defined with $\met > 75$~GeV, therefore implicitly relies on the
assumption that the
\ET \ distributions for photon candidates are similar in both regions.  However, since \MET \ is essentially
a negative vector sum of the \ET \ values of the energy depositions in the calorimeter, it is expected that there
should be a correlation between the value of \MET \ and the \ET \ 
distributions of the physics objects in the event.  This correlation is indeed observed in the TL
control regions: the samples with higher \MET \ values have higher
average photon \ET \ values.
Increasing the minimum \ET \ requirement on the photons to 60~GeV in
the $\MET < 20$~GeV control sample selects events with  kinematic properties 
which are more similar to those of events in the signal region, and therefore the
$\ET > 60$~GeV cut is applied to determine the nominal  $T_{\MET <20 \ {\rm GeV}}$ template shape.
Conservative possible systematic variations of this shape are taken to be the template shapes with \ET \ cuts
 of 50~GeV and 70~GeV on the photon candidates.
The $T_{\MET<20 \ {\rm GeV}}$ template, along with its statistical
and total uncertainties, is shown superimposed on Figure~\ref{figJetZvertTemplates}.

The distribution of arrival times of the photon candidates is used in the analysis
as a cross-check.  Figure~\ref{fig:timingtemplates} shows the timing distribution
expected for selected signal events, for $\Lambda = 120$~TeV and for NLSP lifetime 
values of $\tau = 0.5$~and~$30$~ns. 
The expectations for the backgrounds are
determined using the same data control samples described above. 
It is expected that the performance of the calorimeter timing measurement, as determined using the
second-layer cell with the maximum deposited energy,
should be rather insensitive to the details of the EM shower development. It was verified that
the timing distribution of electrons in $Z\ra ee$ events is very similar to that of loose
photon candidates in the TL control sample with $\MET < 20$~GeV. Therefore, the timing
distribution determined with the larger $Z\ra ee$ sample is characteristic of
the timing performance expected
for all prompt backgrounds, and is shown superimposed on
 Figure~\ref{fig:timingtemplates}.

\begin{figure}[hbpt]
\centering
\includegraphics[width=\columnwidth]{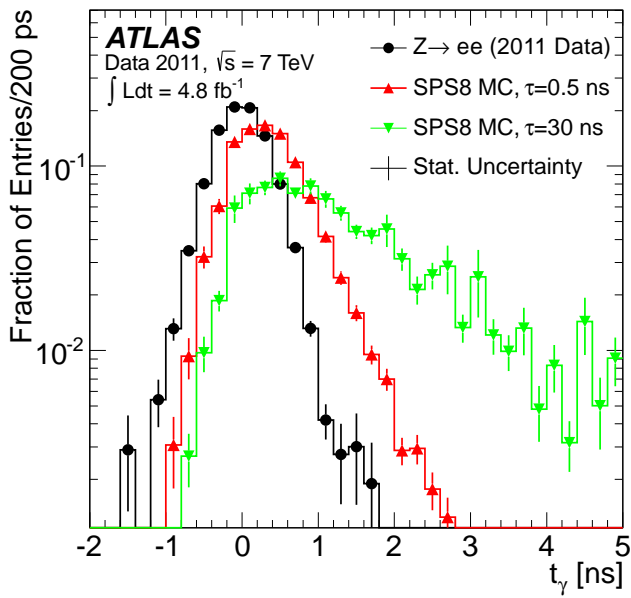}
\caption{The distribution of photon arrival times ($t_\gamma$) expected for SPS8 GMSB signal models with 
$\Lambda = 120$~TeV and for NLSP lifetime 
values of $\tau = 0.5$~and~$30$~ns.
Superimposed is the expectation for prompt
backgrounds, as determined using electrons from $Z\ra ee$ events. The 
uncertainties shown are statistical only.}\label{fig:timingtemplates}
\end{figure}

\section{Systematic Uncertainties}
\label{sec:systematics}

When fitting the photon pointing distribution in data to the templates describing the
expectations from signal and background,
the total number of events is normalized to the 46 events observed in the data in the signal region. In addition,
the background templates are determined using data.  Thus, the analysis
does not rely on predictions of the background normalization or composition, and there are therefore
no systematic uncertainties to consider regarding the normalization of the backgrounds.
As a result, the various systematic uncertainties relevant for the analysis can be divided into two types, 
namely ``flat'' uncertainties that are not a function of $z_{\rm DCA}$ but that affect the overall expected signal
yield, and ``shape'' uncertainties related to the shapes of the unit-normalized signal and background
pointing templates. 

The shape uncertainties, which are correlated across $z_{\rm DCA}$ bins, 
are discussed in Sec.~\ref{sec:templates}; their sizes within each bin
are depicted in Fig.~\ref{figJetZvertTemplates}.
The control region of TL events with \MET \ values in the intermediate range 
from $20-75$~GeV is well described by a fit using the 
$T_{e/\gamma}$ and $T_{\MET < 20 \ {\rm GeV}}$ distributions, providing further 
validation of the background template shapes and their associated systematic uncertainties.

The various flat systematic uncertainties are summarized in Table~\ref{tab:systematics},
including the uncertainty on the integrated luminosity
of $\pm 1.8$\%~\cite{Aad:2013ucp}. The uncertainty on the trigger 
efficiency is $\pm 2.1$\%, which includes a contribution of  $\pm 0.5$\% for the determination of
the diphoton trigger efficiency  using a bootstrap
method~\cite{Aad:2012xs}, and a contribution of $\pm 2$\% taken as a conservative estimate of 
any possible dependence of the trigger efficiency on $z_{\rm DCA}$ or on the delay of signal photons.
Uncertainties on the photon selection,
the photon energy scale, and the detailed material composition of the detector,
as described in Ref.~\cite{2012afa},  result in an uncertainty of $\pm 4.4$\%.  
The uncertainty due to the photon isolation requirement
was estimated by varying the energy leakage
and the pile-up corrections independently, resulting in an
uncertainty of $\pm 1.4$\%. Systematic uncertainties due to the
\MET \ reconstruction, estimated by varying the cluster energies
and the  \MET \ resolution between the measured performance
and MC expectations~\cite{Aad:2012re}, contribute an uncertainty in the range of
$\pm$(1.1--8.2)\%, with the higher uncertainty values applicable 
for lower values of $\Lambda$.  As described in Sec.~\ref{sec:datamc},
variations in the calculated NLO signal cross sections, times the signal acceptance and  
efficiency,  at the level of $\pm$(4.7--6.4)\% occur
when varying the PDFs and factorization and renormalization scales. 
Adding these flat systematic uncertainties in quadrature gives a total
systematic uncertainty on the signal yield in the range of 
$\pm$(7.2--11.7)\%, to which is added a contribution in the range of $\pm$(0.7--5.0)\%
due to statistical uncertainties in the signal MC predictions.

\begin{table}[hbpt]
  \centering
  \begin{tabular}{lc} \hline \hline
Source of Uncertainty & Value   \\ \hline 
Integrated Luminosity & $\pm 1.8$\%    \\ 
Trigger Efficiency   & $\pm 2.1$\%  \\    
Photon ID and \ET \ Scale/Resolution  & $\pm 4.4$\%  \\    
Photon Isolation  & $\pm 1.4$\% \\    
 \MET:\  \ET \ Scale/Resolution  & $\pm$(1.1--8.2)\% \\   
Signal PDF and Scale Uncertainties   & $\pm$(4.7--6.4)\%   \\    \hline 
Total Systematic Uncertainty&  $\  \pm$(7.2--11.7)\% \\ \hline \hline
  \end{tabular}
  \caption{Summary of systematic uncertainties on the total signal yield. The final row
  provides the total of these systematic uncertainties, calculated as the quadrature sum of the various
  contributions.  There is an additional contribution, not shown in the Table, due to
  MC statistics, that ranges between $\pm 0.7$\% and $\pm 5.0$\%, depending mostly
  on the NLSP lifetime.
 }
  \label{tab:systematics}
\end{table}

\section{Results and Interpretation}
\label{sec:results}

The $z_{\rm DCA}$ distribution for the 46 loose photons of the events in the signal region
with $\MET > 75$~GeV is shown in Fig.~\ref{fig:sigfit}.
As expected for SM backgrounds, the distribution is rather narrow, and there is no obvious
sign of a significant excess in the tails that would be expected for GMSB signal photons
originating in the decays of long-lived neutralinos.  There are three events
with $| z_{\rm DCA} | > 200$~mm, including one
with a value of $z_{\rm DCA} = +752$~mm. Some additional information about these 
three events is summarized in Table~\ref{tab:threeevents}.

\begin{figure*}[hbpt]
\centering
\includegraphics[width=\textwidth]{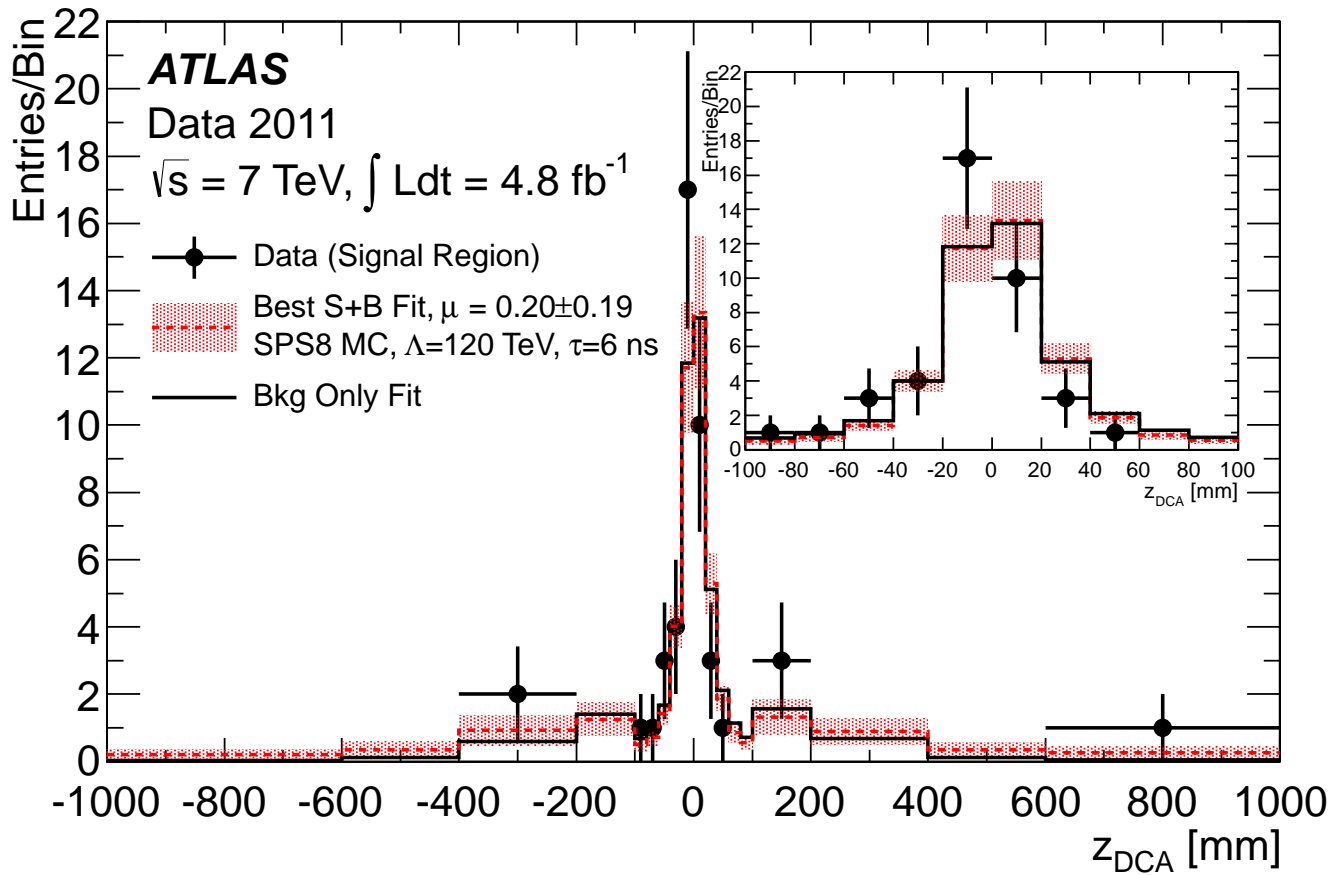}
\caption{The $z_{\rm DCA}$ distribution for the 46 loose photon candidates of the
events in the signal region.
Superimposed are the results of the background-only fit, as well as the
results of the signal-plus-background fit 
for the case of $\Lambda = 120$~TeV and $\tau = 6$~ns. The hatching shows
the total uncertainties in each bin for the signal-plus-background fit, for
which the fitted signal strength 
  is $\mu = 0.20 \pm 0.19$. 
The first (last) bin includes the contribution from underflows (overflows).
The inlay shows an expanded view of the central region, near $z_{\rm DCA} = 0$.}
\label{fig:sigfit}
\end{figure*}

\begin{table*}[hbpt]
\begin{center}
\begin{tabular}{ccccccccc}\hline \hline
Run & Event & \MET & \multicolumn{3}{c}{Loose Photon}  &
\multicolumn{3}{c}{Tight Photon}    \\ \cline{4-9}
Number & Number & (GeV) &  \ET \ (GeV) & $z_{\rm DCA}$ (mm) & \ \ $t_\gamma$ (ps) \ \ &
\ET \ (GeV) & $z_{\rm DCA}$ (mm) & \ \ $t_\gamma$ (ps) \ \ \\ \hline 
 186721& 30399675 &77.1 & 75.9 & \ \ --274 \ \ & 360 & 72.0 & 22 & \ \ 580 \ \ \\ 
 187552& 14929851 &77.3 & 59.4& --262 & 1200 & 87.2 & --120 & 240 \\ 
 191920& 14157929 &77.9 & 56.6& 752 & 2 & 54.2 & 5 & --200 \\ \hline \hline
\end{tabular}
  \caption{Some parameters of the three ``outlier'' events mentioned in the text. }
 \label{tab:threeevents}
 \end{center}
\end{table*}

The timing distribution for the 46 events in the signal region
with $\MET > 75$~GeV is shown in Fig.~\ref{fig:sigtiming}.
The timing distribution is rather narrow, in agreement with the background-only
expectation that is shown superimposed on Fig.~\ref{fig:sigtiming}, and there is no
significant excess in the positive tail that would be expected for GMSB signal photons.  The photon
with the largest time value has $t \approx 1.2$~ns, and is measured
using a calorimeter cell that was read out with medium gain. 
This photon corresponds to one of the three events with $| z_{\rm DCA} | > 200$~mm
that are described in Table~\ref{tab:threeevents}, but not to the most extreme pointing outlier.
For the other two events in Table~\ref{tab:threeevents}, the timing is consistent with
the hypothesis that the photon candidate is in-time.

\begin{figure}[hbpt]
\centering
\includegraphics[width=\columnwidth]{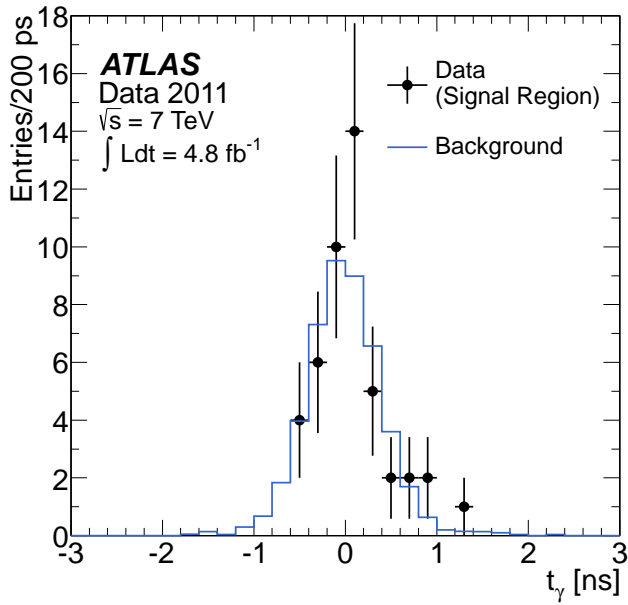}
\caption{The distribution of arrival times ($t_\gamma$) for the 46 loose photon candidates of the
events in the signal region. 
Superimposed for comparison is the shape of the timing distribution expected for background only,
normalized to 46 total events.} 
\label{fig:sigtiming}
\end{figure} 

To determine the final results, the pointing distribution shown in Fig.~\ref{fig:sigfit} is fitted
using the pointing templates described previously. The binning shown in Fig.~\ref{fig:sigfit}
is used in the fit.  The background contribution
is modeled in the fit as a weighted sum of the $T_{e/\gamma}$ and $T_{\MET < 20 \ {\rm GeV}}$ templates.
Background-only fits are performed
to determine the compatibility of the observed pointing distribution with the background-only
hypothesis. Signal-plus-background fits are performed to determine, via profile likelihood fits,
the 95\% CL limit on the signal strength, $\mu$, defined as the number of
fitted signal events divided by the SPS8 expectation for the signal yield. In both cases, the overall normalization is constrained
to the 46 events observed in the signal region in data, and
the  relative weighting of the two background templates is treated as a nuisance parameter in the fit.

\begin{table*}[hbpt]
\begin{center}
\begin{tabular}{ccccccccccc}\hline \hline
Fit & Event & \multicolumn{9}{c}{Range of $| z_{\rm DCA} |$ values  \ \ [mm]} \\ \cline{3-11}
Type & Type & 0--20 & 20--40 & 40--60 & 60--80 & 80--100 & 100--200 & 200--400 & 400--600 & $>$ 600  \\ \hline
- & Data & 27 & 7 & 4 & 1 & 1 & 3 & 2 & 0 & 1\\ \hline 
Bkg Only & Bkg &25.0$\pm$2.2&9.1$\pm$0.8&3.8$\pm$0.3&2.1$\pm$0.5&1.4$\pm$0.4&3.0$\pm$1.1&1.3$\pm$0.5&0.2$\pm$0.1&0.08$\pm$0.03\\ \hline 
Signal & Total &25.1$\pm$4.2&9.3$\pm$1.5&3.3$\pm$0.7&1.6$\pm$0.6&1.1$\pm$0.4&2.6$\pm$1.0&1.8$\pm$0.8&0.7$\pm$0.5&0.5$\pm$0.4\\ 
Plus & Sig &\ \ 0.7$\pm$0.6&0.5$\pm$0.5&0.4$\pm$0.3&0.3$\pm$0.3&0.3$\pm$0.3&1.2$\pm$1.1&1.3$\pm$1.2&0.6$\pm$0.5&0.4$\pm$0.4\\ 
Bkg & Bkg &24.4$\pm$4.2&8.8$\pm$1.5&2.9$\pm$0.8&1.3$\pm$0.7&0.8$\pm$0.6&1.4$\pm$1.5&0.5$\pm$0.7&0.1$\pm$0.1&0.03$\pm$0.04\\ \hline \hline
\end{tabular}
  \caption{Integrals over various
  $|z_{\rm DCA}|$ ranges of the distributions shown in Fig.~\ref{fig:sigfit} for the 46 loose
  photon candidates in the signal region. The numbers of events observed in data are shown, as well as
  the results of a background-only fit and a signal-plus-background fit for the
  case of $\Lambda = 120$~TeV and $\tau = 6$~ns.  The fitted signal strength 
  is $\mu = 0.20 \pm 0.19$. The errors shown correspond to the sum of statistical and
   systematic uncertainties. The numbers of
  signal and background events from the signal-plus-background fit are negatively correlated.}
 \label{tab:fitIntegrals}
 \end{center}
\end{table*}

Table~\ref{tab:fitIntegrals} shows the number of observed events in the various bins of
$z_{\rm DCA}$ shown in Fig.~\ref{fig:sigfit}, except that the number of $z_{\rm DCA}$ bins is
reduced for display purposes by a factor of two by taking the absolute value of $z_{\rm DCA}$.
Included in Table~\ref{tab:fitIntegrals} are the results from the background-only and signal-plus-background
fit, for the case of $\Lambda = 120$~TeV and $\tau = 6$~ns. Comparing the first two rows of
Table~\ref{tab:fitIntegrals} shows that there is fair agreement between the observed
data and the results of the background-only fit, though there is some excess seen in the data for
larger $| z_{\rm DCA} |$ values. The $p_0$ 
value for the background-only hypothesis is $\approx 0.060$, indicating that the slight excess has a
significance equivalent to $\approx 1.5\sigma$.  The excess is dominated by the outlier photon with $z_{\rm DCA} = 752$~mm;
removing this event from the distribution and performing a new fit, one obtains a $p_0$ value of  $\approx 0.30$,
indicating in this case a much better agreement with the background-only model.
   
The last three rows of Table~\ref{tab:fitIntegrals} show the results of the signal-plus-background
fit for the case of $\Lambda = 120$~TeV and $\tau = 6$~ns, including the total number of fitted events in each 
$| z_{\rm DCA} |$ bin, as well as the separate contributions
from signal and from background. 
For this case, the signal-plus-background fit returns a central value for the signal 
strength of $\mu = 0.20 \pm 0.19$, and the fitted value of the nuisance 
parameter specifying the fraction of the background that is attributed to 
the $T_{\MET < 20 \ {\rm GeV}}$ template is $0.32 \pm 0.38$.
The fit results are used to determine, via the CLs method~\cite{Read:2002hq},
95\% CL limits on the number of signal events. The results, derived within the RooStats
framework~\cite{Moneta:2010pm}, are determined for both the
observed limit, where the fit is performed to the pointing distribution of the 46 observed
events in the signal region, and for the expected limit, where the fit is performed to ensembles
of pseudo-experiments generated according to the background-only hypothesis.
The slight excess for larger $| z_{\rm DCA} |$ values 
leads to a result where the observed limit on the number of signal events is somewhat
less restrictive than the expected limit. For the example of $\Lambda = 120$~TeV and 
$\tau = 6$~ns, the observed (expected) 95\% CL limit is 18.3 (9.8) signal events.

By repeating the statistical procedure for various $\Lambda$ and $\tau$
values, the limits are determined as a function of these SPS8 model parameters.
The left plot of Fig.~\ref{fig:1Dlimits} shows the 95\% CL limits on
the number of signal events versus $\tau$, for the case with $\Lambda = 120$~TeV.
The right plot of Fig.~\ref{fig:1Dlimits} shows the 95\% CL limits on the allowed cross section
versus $\tau$, also for the case with $\Lambda = 120$~TeV. Each plot includes a curve indicating the
SPS8 theory prediction for $\Lambda = 120$~TeV. The intersections where the limits cross the
theory prediction show that,  for $\Lambda = 120$~TeV, values of $\tau$ below 8.7~ns are
excluded at 95\% CL, whereas the expected limit would exclude values of $\tau$ below 14.6~ns.

\begin{figure*}[hbpt]
\centering
\includegraphics[width=\columnwidth]{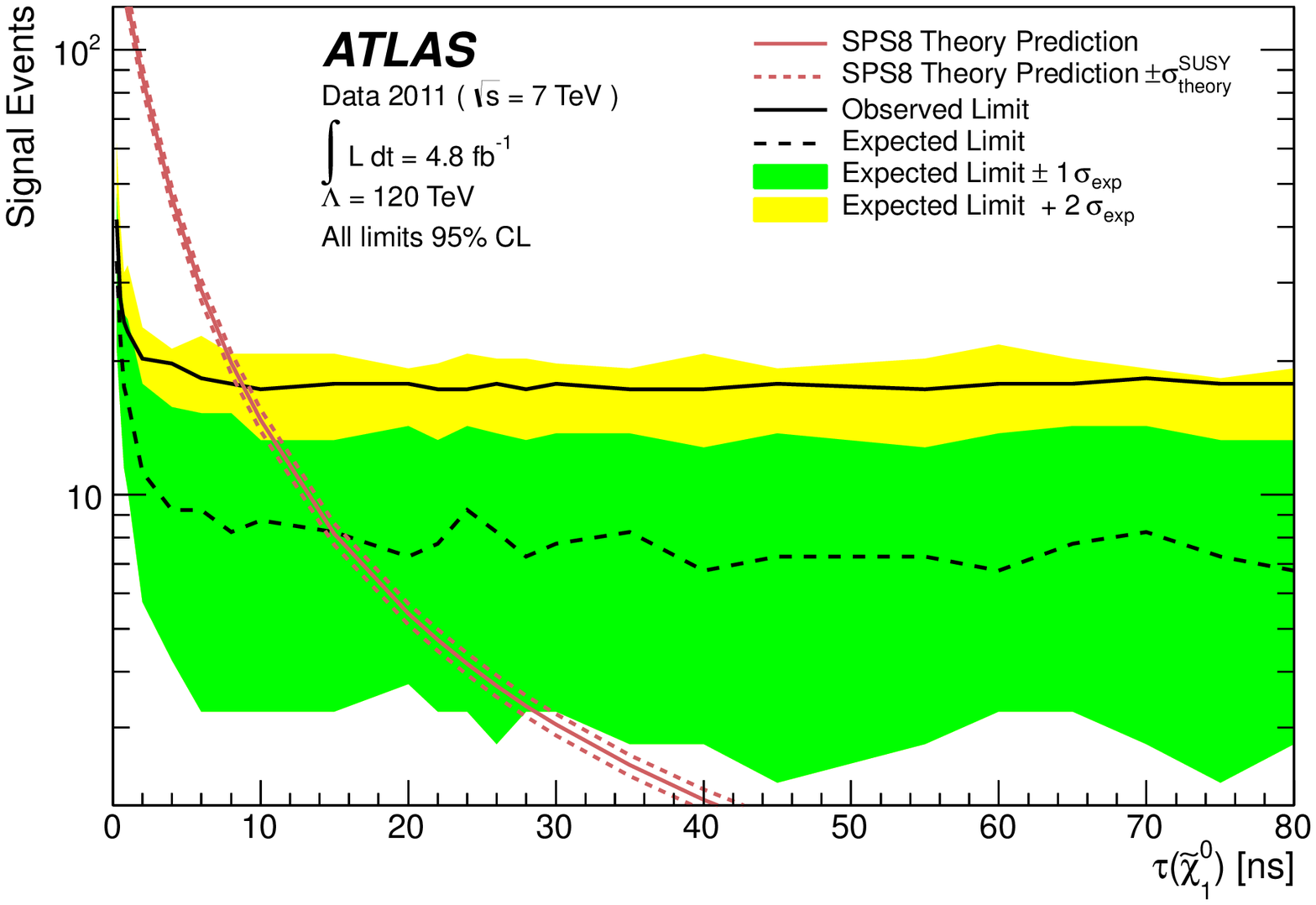}
\includegraphics[width=\columnwidth]{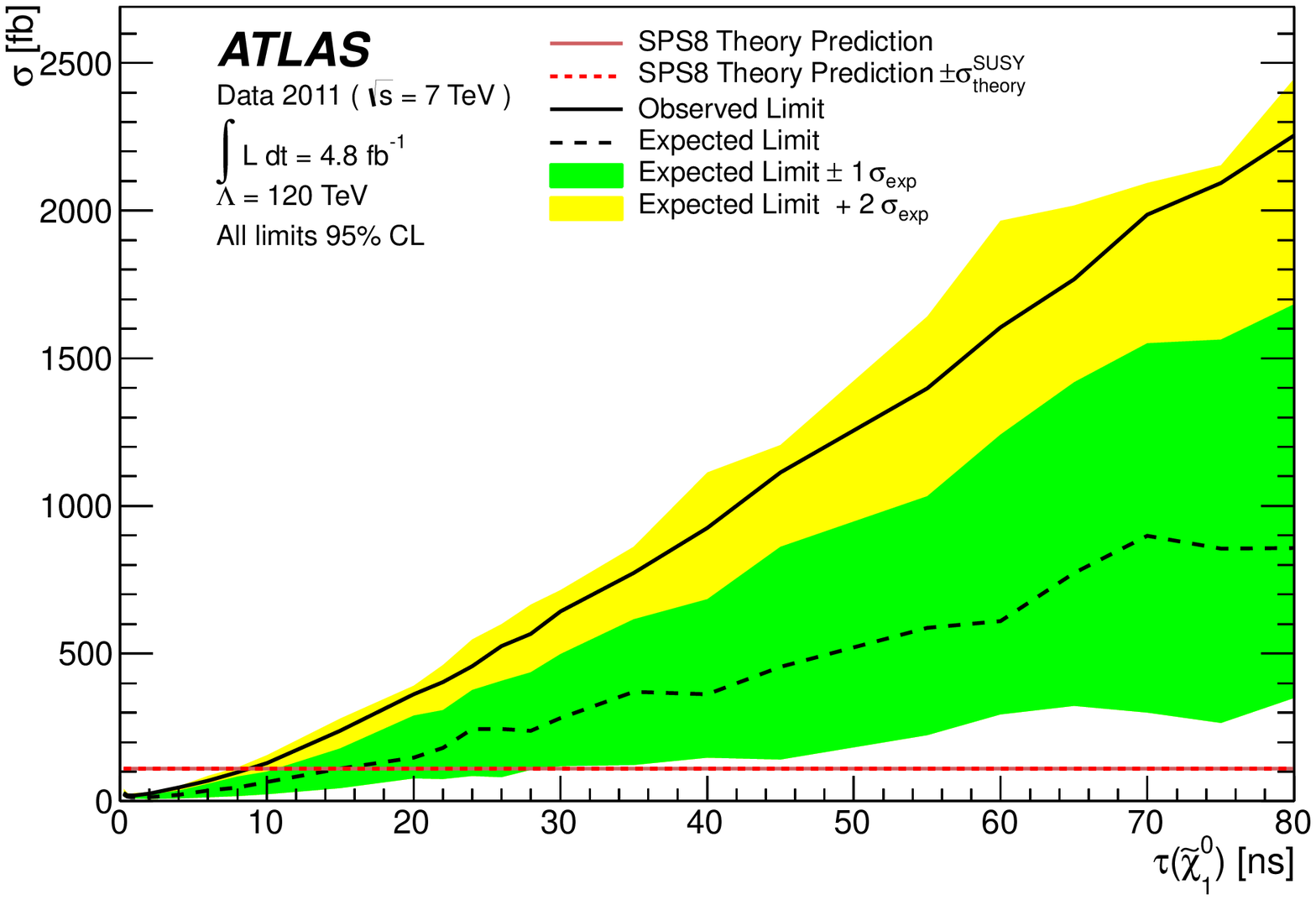}
\caption{95\% CL limits on (left) the number of signal events and
(right) the SPS8 signal cross section, as a function of
NLSP lifetime, for the case of $\Lambda =120$~TeV. The region below the limit
curve is excluded at 95\% CL. }
\label{fig:1Dlimits}
\end{figure*}

Comparing with the theoretical cross section of the
SPS8 GMSB model, the results are converted into an exclusion region in the
two-dimensional plane of $\tau$ versus $\Lambda$, as shown in Fig.~\ref{fig:obslimit}.
Also shown in the figure are corresponding limits on the
lifetime versus the masses of the lightest neutralino and lightest charginos, 
where the relation between $\Lambda$ and sparticle masses is taken from the theory.

\begin{figure*}[hbpt]
\centering
\includegraphics[width=\textwidth]{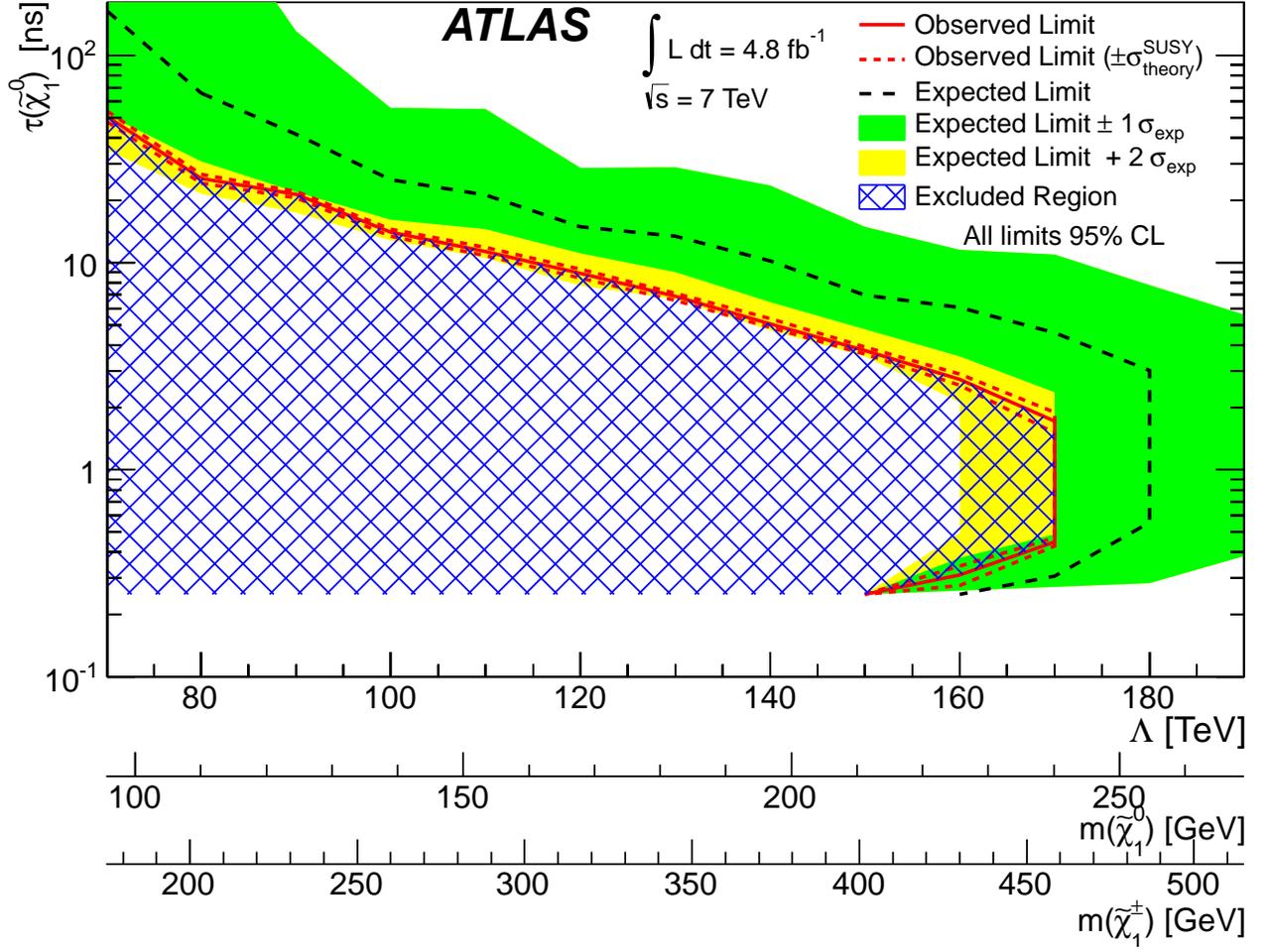}
\caption{The expected and observed limits in the plane of NLSP lifetime versus
$\Lambda$ (and also versus the \ninoone \ or \chinoonepm \ masses), for the SPS8 model. Linear interpolations are shown to connect between
$\Lambda$ values, separated by 10~TeV, for which MC signal samples are available.
The region excluded at 95\% CL is shown as the blue hatched area. The limit is not shown
below an NLSP lifetime of $0.25$~ns, which, due to the MC signal samples available,
is the smallest value considered in the analysis. The green and yellow bands show the $\pm 1\sigma$ and $+2\sigma$ 
variations, respectively, around the expected limit. The $-2\sigma$ variation is not shown since it 
would correspond to a negative number of expected signal events, which is unphysical.Ó
 }
\label{fig:obslimit}
\end{figure*}

\section{Conclusions}
\label{sec:conclusions}

A search has been performed for  non-pointing photons in the
diphoton plus \MET \ final state, using the full data sample of  7 TeV $pp$ collisions recorded in 2011
with the ATLAS detector at the LHC. The analysis uses the  capability of the ATLAS LAr calorimeter 
to measure the flight direction of photons to perform the search, with the precision measurement of
the arrival time of photons used as a cross-check of the results.

No significant evidence for non-pointing photons is observed.
Interpreted in the context of the GMSB SPS8 benchmark model, the results
provide 95\% CL exclusion limits in the plane of $\tau$ (the lifetime of the lightest neutralino) 
versus $\Lambda$ (the effective scale of SUSY breaking) or, alternatively, versus
the mass of the lightest neutralino.  For example, for $\Lambda =$ 70~TeV  (160~TeV), 
NLSP lifetimes between 0.25 and 50.7~ns (2.7~ns) are excluded at 95\% CL.

\input{Acknowledgements}

\bibliographystyle{atlasBibStyleWoTitle}
\bibliography{NPP}

\newpage
\input{atlas_authlist}

\end{document}

%% file: Acknowledgements.tex

\section{Acknowledgements}

We thank CERN for the very successful operation of the LHC, as well as the
support staff from our institutions without whom ATLAS could not be
operated efficiently.

We acknowledge the support of ANPCyT, Argentina; YerPhI, Armenia; ARC,
Australia; BMWF and FWF, Austria; ANAS, Azerbaijan; SSTC, Belarus; CNPq and FAPESP,
Brazil; NSERC, NRC and CFI, Canada; CERN; CONICYT, Chile; CAS, MOST and NSFC,
China; COLCIENCIAS, Colombia; MSMT CR, MPO CR and VSC CR, Czech Republic;
DNRF, DNSRC and Lundbeck Foundation, Denmark; EPLANET, ERC and NSRF, European Union;
IN2P3-CNRS, CEA-DSM/IRFU, France; GNSF, Georgia; BMBF, DFG, HGF, MPG and AvH
Foundation, Germany; GSRT and NSRF, Greece; ISF, MINERVA, GIF, DIP and Benoziyo Center,
Israel; INFN, Italy; MEXT and JSPS, Japan; CNRST, Morocco; FOM and NWO,
Netherlands; BRF and RCN, Norway; MNiSW, Poland; GRICES and FCT, Portugal; MERYS
(MECTS), Romania; MES of Russia and ROSATOM, Russian Federation; JINR; MSTD,
Serbia; MSSR, Slovakia; ARRS and MIZ\v{S}, Slovenia; DST/NRF, South Africa;
MICINN, Spain; SRC and Wallenberg Foundation, Sweden; SER, SNSF and Cantons of
Bern and Geneva, Switzerland; NSC, Taiwan; TAEK, Turkey; STFC, the Royal
Society and Leverhulme Trust, United Kingdom; DOE and NSF, United States of
America.

The crucial computing support from all WLCG partners is acknowledged
gratefully, in particular from CERN and the ATLAS Tier-1 facilities at
TRIUMF (Canada), NDGF (Denmark, Norway, Sweden), CC-IN2P3 (France),
KIT/GridKA (Germany), INFN-CNAF (Italy), NL-T1 (Netherlands), PIC (Spain),
ASGC (Taiwan), RAL (UK) and BNL (USA) and in the Tier-2 facilities
worldwide.

%% file: atlas_authlist.tex
\begin{flushleft}
{\Large The ATLAS Collaboration}

\bigskip

G.~Aad$^{\rm 48}$,
T.~Abajyan$^{\rm 21}$,
B.~Abbott$^{\rm 112}$,
J.~Abdallah$^{\rm 12}$,
S.~Abdel~Khalek$^{\rm 116}$,
A.A.~Abdelalim$^{\rm 49}$,
O.~Abdinov$^{\rm 11}$,
R.~Aben$^{\rm 106}$,
B.~Abi$^{\rm 113}$,
M.~Abolins$^{\rm 89}$
% ...
\bigskip

% ...
$^{11}$ Institute of Physics, Azerbaijan Academy of Sciences, Baku, Azerbaijan\\
$^{12}$ Institut de F\'isica d'Altes Energies and Departament de F\'isica de la Universitat Aut\`onoma de Barcelona and ICREA, Barcelona, Spain\\
% ...
$^{21}$ Physikalisches Institut, University of Bonn, Bonn, Germany\\
% ...
$^{48}$ Fakult\"{a}t f\"{u}r Mathematik und Physik, Albert-Ludwigs-Universit\"{a}t, Freiburg, Germany\\
$^{49}$ Section de Physique, Universit\'e de Gen\`eve, Geneva, Switzerland\\
% ...
$^{89}$ Department of Physics and Astronomy, Michigan State University, East Lansing MI, United States of America\\
% ...
$^{106}$ Nikhef National Institute for Subatomic Physics and University of Amsterdam, Amsterdam, Netherlands\\
% ...
$^{112}$ Homer L. Dodge Department of Physics and Astronomy, University of Oklahoma, Norman OK, United States of America\\
$^{113}$ Department of Physics, Oklahoma State University, Stillwater OK, United States of America\\
% ...
$^{116}$ LAL, Universit\'e Paris-Sud and CNRS/IN2P3, Orsay, France\\
% ...
\end{flushleft}
